\documentclass[preprint,prl,aps,superscriptaddress]{revtex4-1}
\usepackage{graphicx}
\usepackage{bm}
\usepackage{amssymb,amsmath}
\usepackage{color}
\usepackage{float}

\begin{document}

\title{Determination of the quantized topological magneto-electric effect in
topological insulators from Rayleigh scattering}
\author{Lixin Ge}
\author{Tianrong Zhan}
\altaffiliation[Present address: ]{Bartol Research Institute, University of Delaware, Newark,
Delaware 19716, USA}
\affiliation{Department of Physics, Key laboratory of Micro and Nano Photonic Structures
(MOE), and Key Laboratory of Surface Physics, Fudan University, Shanghai 200433, China}
\author{Dezhuan Han}
\email{dzhan@cqu.edu.cn}
\affiliation{Department of Applied Physics, College of Physics,
Chongqing University, Chongqing 400044, China}
\affiliation{Department of Physics, Key laboratory of Micro and Nano Photonic Structures
(MOE), and Key Laboratory of Surface Physics, Fudan University, Shanghai 200433, China}
\author{Xiaohan Liu}
\author{Jian Zi}
\email{jzi@fudan.edu.cn}
\affiliation{Department of Physics, Key laboratory of Micro and Nano Photonic Structures
(MOE), and Key Laboratory of Surface Physics, Fudan University, Shanghai 200433, China}
\date{\today }

\begin{abstract}
Topological insulators (TIs) exhibit many exotic properties. In particular, a
topological magneto-electric (TME) effect, quantized in units of the fine
structure constant, exists in TIs. In this Letter, we study theoretically
the scattering properties of electromagnetic waves by TI circular cylinders
particularly in the Rayleigh scattering limit. Compared with ordinary dielectric
cylinders, the scattering by TI cylinders shows many unusual features
due to the TME effect. Two proposals are suggested to determine
the TME effect of TIs simply based on measuring the electric-field
components of scattered waves in the far field at one or two scattering angles.
Our results could also offer a way to measure the fine structure constant.
\end{abstract}

\maketitle

\section{Introduction}

The scattering of electromagnetic (EM) waves by small particles is a common
optical phenomenon \cite{boh:83,tsa:00}. According to the size of scatters,
it can be classified into Rayleigh (for scatterer sizes much smaller
than the wavelength) and Mie scattering (for scatterer sizes comparable to the
wavelength). The most known example of Rayleigh scattering is the blue color of
the sky wherein the scattering intensity by the gas molecules in the
atmosphere varies inversely with the fourth power of the wavelength.
The scattering of EM waves depends not only on the scatterer size and geometry
but also on the EM properties of scatters. In addition to conventional dielectric
ones, scatterers made of emerging artificial materials have received broad
interest in recent years and many extraordinary scattering properties
have been revealed \cite{pen:06, rua:10,lai:09,ker:83,tho:12}.
For instance, the scattering of an object composed of metamaterial
in the far field could be reduced to zero (cloaking) \cite{pen:06},
or even transformed to other objects of different appearance \cite{lai:09}.
For a subwavelength
nanorod consisting of multiple concentric layers of dielectric and plasmonic
materials, its scattering cross-section can far exceed the single-channel limit,
leading to superscattering \cite{rua:10}.
For particles with magnetic responses, many unusual EM scattering properties
such as zero forward scattering has been proposed and confirmed
experimentally \cite{ker:83}. Even being one-atom thick, arrays of
graphene nano-disks could nearly completely absorb infrared light at certain
resonant wavelengths \cite{tho:12}.

As an emerging phase in condensed matter physics, topological insulators (TIs)
has been of great interest in recent years \cite{has:10,qi:11,she:13} due to their
exotic properties. In the bulk, TIs resemble ordinary insulators
possessing a bulk energy gap. However, their surface states are gapless (metallic)
and protected topologically by time-reversal symmetry. TI materials have been
theoretically predicted and experimentally confirmed in several material systems
\cite{ber:06,fu:07}. In additional to exotic electronic and transport properties,
TIs show many unusual EM properties \cite{qi:09,tse:10,mac:10,rag:10,gru:11}.
For instance, a point charge atop the surface of a TI can induce an image
magnetic monopole \cite{qi:09}. In a TI thin film, there exist a giant magneto-optical
Kerr effect and an interesting Faraday effect with a universal rotation angle
defined by the fine structure constant \cite{tse:10}. From the Kerr and Faraday
angles in a TI thick film, one can even determine the half-quantized Hall
conductance of the two TI surfaces independently without knowing the material
details \cite{mac:10}. On TI surfaces, the surface plasmon modes can
even couple to spin waves, forming interesting hybridized spin-plasmon
modes \cite{rag:10}. In two TI plates, Casimir forces could even be
switched to be repulsive \cite{gru:11} although they are usually attractive
in two ordinary dielectric plates. All these interesting phenomena
stem from the topological magneto-electric (TME) effect arising from
the unusual EM response in TIs \cite{qi:08}.

In this Letter, we study theoretically the scattering of EM waves by
circular TI cylinders, particularly in the Rayleigh scattering limit.
Unusual scattering properties due to the TME effect are revealed.
In three-dimensional TIs, the EM response can be described by a
Lagrangian consisting of a conventional Maxwell term
and an additional term related to the TME effect \cite{qi:08},
$\Delta{\cal{L}}=(\theta\alpha/4\pi^{2})\mathbf{E}\cdot\mathbf{B}$,
where $\mathbf{E}$ and $\mathbf{B}$ are the electric and magnetic
fields, respectively;
$\alpha= e^{2}/\hbar c$ is the fine structure constant; and
$\theta=(2p+1)\pi$ with $p$ being an integer is a quantized angular
variable to characterize the TME effect, known as the axion angle in
particle physics \cite{wil:87}. By breaking the time-reversal
symmetry at the surface (e.g., coating a TI with a thin ferromagnetic layer),
a surface energy gap will open up so that the value of $\theta$ can
be specified definitely as discussed in Refs. \onlinecite{qi:08,ess:09}.
Indeed, $\theta$ gives a half-quantized Hall conductance
$\sigma _{xy}=(p+1/2)e^{2}/{h}$ which can be viewed as the
origin of the TME effect \cite{qi:08}. In TIs, the propagation of EM waves
can still be described by conventional Maxwell's equations. However,
owing to the presence of the TME effect the constitutive relations in TIs
should be modified as \cite{qi:08},
$\mathbf{D} =\mathbf{\varepsilon E}-\overline{\alpha } \mathbf{B}$
and
$\mathbf{H} =\mathbf{B}/\mu +\overline{\alpha } \mathbf{E}$,
where $\mathbf{D}$ and $\mathbf{H}$ are respectively the electric
displacement and the magnetic field strength,
$\varepsilon$ and $\mu$ are respectively the dielectric constant and
magnetic permeability, and $\overline{\alpha }=(\theta/\pi)\alpha$
is a quantized quantity in units of the fine structure constant. Note that
the effective description of the modified constitutive relations of TIs
applies only for the photon energy $\hbar\omega$ much smaller
than both the bulk and surface energy gaps, where $\omega$ is the
angular frequency of EM waves.
Compared to conventional media such as anisotropic ones \cite{boh:78},
the scattering of EM waves by a TI cylinder differs in an additional contribution
resulting from the TME effect, basically a surface and topological effect that
gives rise to many unique and novel quantum phenomena
\cite{qi:09,tse:10,mac:10,rag:10,gru:11}.

\section{Scattering of EM waves by a TI cylinder}

\begin{figure}[H]
\centerline{\includegraphics[width=7.cm]{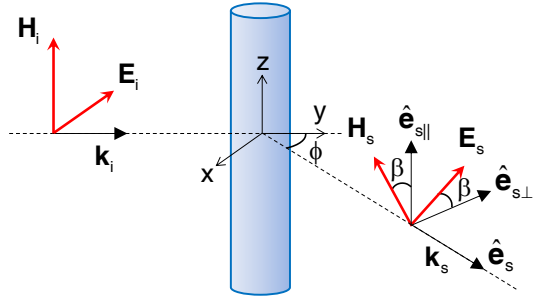}}
\caption{(color online). Schematic view of a circular TI cylinder
placed along the $z$ axis. The subscripts $i$ and $s$ denote incident
and scattered waves, respectively. TE waves with a wave vector
$\mathbf{k}_{i}=k\hat{\mathbf{y}}$ are incident perpendicularly
to the TI cylinder, where
$k=\sqrt{\varepsilon_{\text{b}}\mu_{\text{b}}}\omega/c$.
Another coordinate system is introduced
to describe scattered waves in the far field with the orthonormal
basis vectors:
$\hat{\mathbf{e}}_{s}=\mathbf{k}_{s}/k$ ($\mathbf{k}_{s}$
is the wave vector along the wave normal of scattered waves and
lies in the $x$-$y$ plane),
$\hat{\mathbf{e}}_{s\perp}$ which lies in the $x$-$y$ plane
but perpendicular to both $\hat{\mathbf{e}}_{s}$ and
$\hat{\mathbf{e}}_{s||}$,
and $\hat{\mathbf{e}}_{s||}=\hat{\mathbf{z}}$.
Note that $\mathbf{k}_{s}$ makes a scattering angle $\phi$
with the $y$ axis.
Both the electric and magnetic fields of scattered waves lie in the
$\hat{e}_{s\perp}$-$\hat{e}_{s||}$ plane but
may rotate by the same angle $\beta$ due to the TME effect.}
\label{fig1}
\end{figure}

The system under study is schematically shown in Fig.~\ref{fig1}.
A circular TI cylinder with a radius $r$ is placed along the $z$ axis.
In this study, we focus on transverse electric (TE) waves (with the
magnetic field along the TI cylinder) which are incident perpendicularly
to the TI cylinder. Transverse magnetic (TM) incident waves
(with the electric field along the TI cylinder) can be discussed similarly.
The dielectric constant and magnetic permeability of the TI are
denoted by $\varepsilon$ and $\mu$, respectively, and those of the background
are $\varepsilon _{\text{b}}$ and $\mu _{\text{b}}$. The axion angle of
the TI is $\theta=(2p+1)\pi$ while the background takes a trivial axion angle
$\theta=0$ for simplicity. To break the time-reversal symmetry on the
surface, the TI cylinder is coated with an ultrathin ferromagnetic layer which
plays no role in the scattering since its thickness is much smaller than both
the radius of the TI cylinder and the wavelength of EM waves considered
\cite{wei:13}.

Based on the standard multipole expansion theory \cite{boh:83,tsa:00}, we can 
solve the scattering problem of EM waves by a circular TI cylinder with the 
modified constitutive relations and the conventional boundary conditions at the
boundary between the TI cylinder and the background. The scattering
coefficients \{$a_{n}$\} and \{$b_{n}$\}, related respectively to the
electric and magnetic multipoles of order $n$, can be obtained.
With these scattering coefficients, scattering properties of the TI
cylinder can be obtained accordingly. It can be verified
that $a_{-n}=a_{n}$ and $b_{-n}=b_{n}$, similar to those in
ordinary dielectric cylinders \cite{boh:83,tsa:00}. It should be
mentioned that \{$a_{n}$\} and \{$b_{n}$\}
are polarization-dependent. In other words, there exist two
independent sets of the scattering coefficients,
\{$a_{n,\text{TE}}$,\ $b_{n,\text{TE}}$\} and
\{$a_{n,\text{TM}}$,\ $b_{n,\text{TM}}$\}. For an incident wave
with an arbitrary polarization, its scattering properties can be
discussed since it can be decomposed as a linear combination of TE
and TM waves.

\begin{figure}[H]
\centerline{\includegraphics[width=8.5cm]{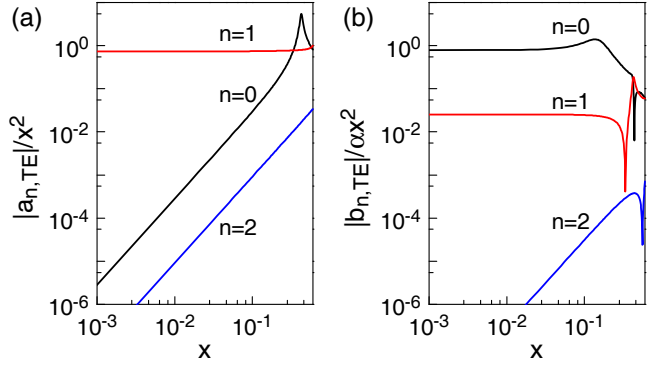}}
\caption{(color online). Scattering coefficients of TE incident waves for
a TI cylinder in air ($\varepsilon_{\text{b}}=\mu_{\text{b}}=1$)
with $\varepsilon=30$ and $\mu=1$.
The axion angle $\theta=\pi$ is taken for the TI. Note that
in (a) $|a_{n,\text{TE}}|$ is normalized by $x^{2}$ and in (b)
$|b_{n,\text{TE}}|$
is normalized by $\alpha x^{2}$.}
\label{fig2}
\end{figure}

Compared with ordinary dielectric cylinders, extra contributions
resulting from the TME effect appear in both \{$a_{n}$\} and \{$b_{n}$\},
leading to many unusual scattering properties. For example,
for an ordinary dielectric cylinder a TE incident wave cannot
excite the magnetic multipoles because of $b_{n,\text{TE}}=0$
(not valid for $b_{n,\text{TM}}$ in general).
However, for a TI cylinder $b_{n,\text{TE}}$ does
not vanish in general, implying that the magnetic multipoles can be
excited. The underlying physics lies in the TME effect whereby an
electric (magnetic) field can induce a magnetic (electric) polarization.
In Fig. \ref{fig2}, the scattering coefficients of a TI cylinder
for TE incident waves as a function of the size parameter $x=kr$ is shown.
In general, the electric multipoles give much larger
contributions to the scattering than the magnetic multipoles. In the Rayleigh
scattering limit ($x\ll 1$ and $mx\ll 1$ with
$m=\sqrt{\varepsilon\mu/\varepsilon_{\text{b}}\mu_{\text{b}}}$),
for TE incident waves it can be shown that only the following scattering
coefficients have order of $x^{2}$,
\begin{subequations}
\begin{eqnarray}
a_{1,\text{TE}}&=& -\frac{i\pi (2m^{2}-2+\overline{\alpha }^{2})}
{4(2m^{2}+2+\overline{\alpha }^{2})}x^{2}+O(x^{4}),\\
b_{0,\text{TE}}&=& \frac{i\pi \overline{\alpha }}{4}x^{2}+O(x^{4}),\\
b_{1,\text{TE}}&=& -\frac{i\pi \overline{\alpha }}{2(2m^{2}+2+
\overline{\alpha }^{2})}x^{2}+O(x^{4}).
\end{eqnarray}
\end{subequations}
All other scattering coefficients have order of $x^{4}$ or
higher, and can be hence neglected in Rayleigh scattering.
In other words, in Rayleigh scattering only the electric dipole, and
magnetic monopole and dipole play roles for TE incident waves.
In the Mie scattering regime ($x\sim 1$), however, both the electric and
magnetic multipoles will contribute to the scattering. Interestingly,
at certain frequencies resonant peaks appear, known as Mie
resonances \cite{boh:83,tsa:00}, corresponding to the resonant excitations
of the electric or magnetic multipoles. For example,
the peak at $x=0.425$ in $|a_{0,\text{TE}}|$ corresponds to the
resonant excitation of the electric monopole while the peak at $x=0.135$
in $|b_{0,\text{TE}}|$ implies
the resonant excitation of the magnetic monopole. In addition to resonant
peaks, there exist sharp dips in $|b_{n,\text{TE}}|$ showing
a \textit{anti-resonance} behavior. The underlying physics is that
at the dips a TE incident wave cannot induce a transverse magnetic field,
leading to $b_{n,\text{TE}}=0$, i.e., the absence of the TME
effect at the dips. For TM incident waves on the other side, it can be
shown that only the scattering coefficients
$a_{0,\text{TM}}$, $a_{1,\text{TM}}$, $b_{0,\text{TM}}$,
and $b_{1,\text{TM}}$ have the $O(x^{2})$ term, and all other coefficients are
in order $x^{4}$ or higher, namely, $a_{0,\text{TM}}=b_{0,\text{TE}}$,
$a_{1,\text{TM}}=b_{1,\text{TE}}$,
$b_{0,\text{TM}}=-i\pi \left(m^{2}-1+
\overline{\alpha }^{2}\right)x^{2}/4+O(x^{4})$,
and $b_{1,\text{TM}}=-\overline{\alpha }b_{1,\text{TE}}/2$.
Different from TE incident waves, a TM incident wave can excite
the electric monopole in Rayleigh scattering.

To obtain the fields of scattered waves in the far field, an amplitude
scattering matrix $T$ is usually introduced which relates the
electric field of scattered waves to that of incident waves
\cite{boh:83,tsa:00}
\begin{equation}
\left(
\begin{array}{c}
E_{s||} \\
E_{s\perp}
\end{array}
\right) =e^{i3\pi /4}\sqrt{\frac{2}{\pi k\rho}}
e^{ik\rho}\left(
\begin{array}{cc}
T_{1} & T_{4} \\
T_{3} & T_{2}
\end{array}
\right) \binom{E_{i||}}{E_{i\perp}},
\label{sca-mat}
\end{equation}
where $E_{||}$ and $E_{\perp}$ are the components of the electric
field parallel and perpendicular to the TI cylinder, respectively;
and $\rho$ is the radial distance from the center of the
TI cylinder in the $x$-$y$ plane. From the definitions in Fig. \ref{fig1},
$E_{i||}=\mathbf{E}_{i}\cdot \hat{\mathbf{z}}$,
$E_{i\perp}=-\mathbf{E}_{i}\cdot \hat{\mathbf{x}}$,
$E_{s||}=\mathbf{E}_{s}\cdot \hat{\mathbf{e}}_{s||}$, and
$E_{s\perp}=\mathbf{E}_{s}\cdot \hat{\mathbf{e}}_{s\perp}$.
The elements of the amplitude scattering matrix $T$ are given by
$T_{1} =\sum_{n=-\infty}^{\infty}e^{-in\phi}b_{n,\text{TM}}$,
$T_{2} =\sum_{n=-\infty}^{\infty}e^{-in\phi}a_{n,\text{TE}}$,
$T_{3} =\sum_{n=-\infty}^{\infty}e^{-in\phi}a_{n,\text{TM}}$,
and $T_{4} =\sum_{n=-\infty}^{\infty}e^{-in\phi}b_{n,\text{TE}}$.
Obviously, the elements $T_{2}$ and $T_{3}$ are associated with the
electric multipoles while $T_{1}$ and $T_{4}$ are related to the magnetic
multipoles. In ordinary dielectric cylinders, the condition
$b_{n,\text{TE}}=0$ for TE incident waves leads to that
the scattering matrix $T$ should be diagonal, i.e.,
$T_{3}=T_{4}=0$. In TI cylinders, however, $T$
is not diagonal in general since $b_{n,\text{TE}}$ may not be zero
owing to the TME effect.

\begin{figure}[H]
\centerline{\includegraphics[width=7.cm]{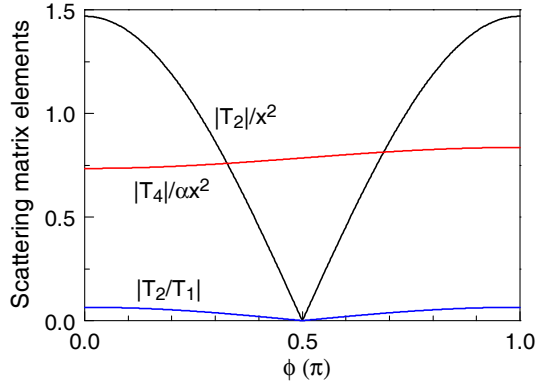}}
\caption{(color online). Scattering matrix elements in the Rayleigh
scattering limit as a function of the scattering angle $\phi$ for a TI cylinder
in air ($\varepsilon_{\text{b}}=\mu_{\text{b}}=1$) with
$\varepsilon=30$ and $\mu=1$.
The axion angle is taken to be $\theta=\pi$. Note that $|T_{2}|$ is
normalized by $x^{2}$ and $|T_{4}|$
is normalized by $\alpha x^{2}$.}
\label{fig3}
\end{figure}

In the Rayleigh scattering limit, it can be shown that the scattering matrix
elements $T_{i}$ can be simplified to the following forms
\begin{subequations}
\begin{eqnarray}
T_{1} &=&-i\pi x^{2}\left[ \frac{m^{2}-1+\overline{\alpha }^{2}}{4}-
\frac{\overline{\alpha }^{2}\cos \phi}{2\left(2m^{2}+2+\overline{\alpha }^{2}\right)}
\right],  \\
T_{2} &=&-i\pi x^{2}\frac{\left(2m^{2}-2+\overline{\alpha }^{2}\right)\cos \phi}
{2\left(2m^{2}+2+\overline{\alpha }^{2}\right)},
\\
T_{3} &=&T_{4}=i\pi x^{2}\left( \frac{\overline{\alpha }}{4}-\frac{
\overline{\alpha }\cos \phi}{2m^{2}+2+\overline{\alpha }^{2}}\right).
\end{eqnarray}
\end{subequations}
Obviously, the scattering matrix elements $T_{i}$ depends not only on the
scattering angle $\phi$ but also on the axion angle. In Fig. \ref{fig3},
the scattering matrix elements as a function of $\phi $ are shown.
The element $|T_{2}|$ depends strongly on $\phi$ showing a
linear relation with $\cos \phi$ in Rayleigh scattering.
As a result, $|T_{2}|$ has two maxima at $\phi=0$ and $\pi$, corresponding to
the forward and backward scattering, respectively. It vanishes at $\phi =\pi /2$.
By analyzing the origin, the element has a dominant contribution from
$a_{1,\text{TE}}$, namely, the electric dipole. In contrast, the element $|T_{4}|$,
which is linearly proportional to $\overline{\alpha }$ in Rayleigh scattering, shows
a rather weak dependence on $\phi $ since there are two contributions from
$b_{0,\text{TE}}$ and $b_{1,\text{TE}}$, associated with the magnetic
monopole and dipole, respectively.

From the amplitude scattering matrix, information on scattering properties
can be inferred. For instance, for a TE incident wave
the scattered waves in general are still dominantly TE-polarized for the
scattering angle away from $\phi=\pi/2$,
although there is a very small rotation of the polarization due to the TME effect.
This is because $|T_{4}|$ is much smaller than $|T_{2}|$ except for $\phi$
in the vicinity of $\pi/2$. Around $\phi=\pi/2$, the polarization of the
scattered waves undergoes a drastic change, and
the scattered wave becomes purely TM-polarized at $\phi=\pi/2$. This
is a strong manifestation of the TME effect. For a TM incident wave,
however, the scattered wave is always dominantly TM-polarized even at $\phi=\pi/2$.
This is why in this study we focus on TE rather than TM incident waves.
From the scattering matrix, the measure of the TME effect is associated with the
ratio $|T_{4}/T_{2}|$ ($|T_{3}/T_{1}|$) for TE (TM) incident waves. Note
that $|T_{1}|$ is much larger than both $|T_{2}|$ and $|T_{3}|$ for any
$\phi$. Thus, the ratio $|T_{3}/T_{1}|$, basically a manifestation of the TME effect,
is more than an order of magnitude smaller than $|T_{4}/T_{2}|$ which is related to TE
incident waves.

\section{Determination of the TME effect}

To explore a quantitative determination of the TME effect, we can conduct
Rayleigh scattering experiments with TE incident waves as schematically
shown in Fig. \ref{fig1}. In the far field, both electric-field components of
scattered waves $|E_{s||}|$ and $|E_{s\perp}|$ are measurable quantities
and we can thus define a measurable quantity $R(\phi )= |E_{s||}/E_{s\perp}|$.
For TE incident waves in the Rayleigh scattering limit,
$R(\phi )=|T_{4}/T_{2}|
=\overline{\alpha}[d-(d-1)\cos \phi]/(2|\cos\phi|)$,
where $d=(2m^{2}+2+\overline
{\alpha }^{2})/(2m^{2}-2+\overline{\alpha }^{2})$ is weakly
dependent on $\overline{\alpha }$ since $\overline{\alpha }\ll m$
and is basically a bulk parameter. By neglecting the
$\overline{\alpha }^{2}$ terms
in $d$, we can obtain
\begin{equation}
\overline{\alpha }\simeq \frac{2 R(\phi )\left(m^{2}-1\right)|\cos\phi| }
{m^{2}+1-2\cos\phi}.
\label{mea11}
\end{equation}
This offers a simply way to determine $\overline{\alpha }$
if $R(\phi )$ is measured at a certain scattering angle $\phi$ and the material
parameter $m$ is known. Note that $\overline{\alpha }$
cannot be determined at $\phi=\pi/2$ from Eq. (\ref{mea11})
since $T_{2}$=0. Thus, a scattering angle $\phi\neq\pi/2$ should
be chosen if using Eq. (\ref{mea11}).
From Eq. (\ref{sca-mat}) and $T_{4}$, however, $\overline{\alpha }$
can still be determined at $\phi=\pi/2$ as
\begin{equation}
\overline{\alpha}=\frac{2}{x^2}\sqrt{\frac{2k\rho}{\pi}}
\left|\frac{E_{s||}}{E_{i\perp}}\right|,
\label{mea12}
\end{equation}
provided that $|E_{s||}/E_{i\perp}|$ is measured, and the radius of the
TI cylinder, the wavelength of the TE incident waves,
and the distance between the TI cylinder and the detector are
known.

Obviously, the one-angle measurement of $\overline{\alpha }$
based on either Eq. (\ref{mea11}) or (\ref{mea12}) is dependent
on material parameters such as $m$ or $x$.
To achieve a determination that is independent of material parameters,
we can do the measurement twice at two different scattering angles $\phi_{1}$
and $\phi_{2} $. With the two observables $R(\phi _{1})$ and
$R(\phi _{2})$ and eliminating the material-dependent
quantity $d$, $\overline{\alpha }$ can be expressed as
\begin{equation}
\overline{\alpha}=\frac{2}{1-\kappa }
\left[R(\phi _{1})\text{sgn}(\cos\phi_{1})-
\kappa R(\phi _{2})\text{sgn}(\cos\phi_{2})\right],
\label{mea2}
\end{equation}
where $\kappa =\cos\phi _{2}(1-\cos \phi _{1})/
[\cos \phi _{1}(1-\cos \phi _{2})]$ is a parameter depending only on
the scattering angles. The to-be-measured $\overline{\alpha }$ is now
only a function of
$\phi _{1,2}$ and $R(\phi_{1,2})$. As in the one-angle measurement
based on Eq. (\ref{mea11}), the scattering angle of $\pi/2$ should be
avoided.

Equations (\ref{mea11})-(\ref{mea2}) are the most important
results in this study. Although the one-angle measurement based on
Eq. (\ref{mea11}) or (\ref{mea12}) is simple, it is, however,
dependent on material parameters since we have to know the material
parameter $m$ at the frequency of the incident waves or the radius of
the TI cylinder. In contrast, the two-angle measurement based on
Eq. (\ref{mea2}) is material-independent. It needs only the measured
quantity $R(\phi)$ at two scattering angles.
Importantly, this two-angle
measurement can still work even for TIs with bulk free carriers. This is
very meaningful since the present TI materials are still non-insulating
in the bulk.

From $\overline{\alpha }$,
the axion angle $\theta$ could be directly inferred, from which the
half-quantized Hall conductance of the surface of the TI cylinder
can be obtained. From the obtained $\overline{\alpha }$, it also offers a
way to measure the fine structure constant $\alpha$ since the axion
angle is quantized. Practically, such Rayleigh scattering experiments can
be conducted in the microwave regime. The radius of TI cylinders should be
of the order of micrometers or tens of micrometers which well satisfies
the Rayleigh-scattering-limit condition. To reduce the influence of incident
waves, scattering angles around $\pi/2$ are suggested.

\section{Conclusion}

In summary, we study Rayleigh scattering of EM waves by circular TI
cylinders based on a multipole expansion theory.
Based on the unconventional scattering features, two proposals
are suggested to measure the quantized TME effect. The
two-angle has a promising feature of material-independence.
Our proposal offers a way to determine the axion angle or a method
to measure the fine structure constant.

\section*{Acknowledgements}

We acknowledge helpful discussions with S.-Q. Shen. This work is
supported by the 973 Program (Grant Nos. 2013CB632701 and
2011CB922004). The research of H.D.Z., X.H.L. and J.Z. is further supported
by the National Natural Science Foundation of China.

\section{}
\begin{appendix}

\setcounter{equation}{0}

\subsection{Appendix A: Multipole expansion and Rayleigh scattering}
\renewcommand{\theequation}{A.\arabic{equation}}

For EM waves incident perpendicularly to an infinite circular cylinder,
the scattering can be treated analytically by the standard multipole expansion
theory \cite{boh:83,tsa:00}. Under the framework of the theory, in the cylindrical
coordinate system all associated waves can be expanded by the following vector
cylindrical harmonics
\begin{subequations}
\begin{eqnarray}
\mathbf{M}_{n}^{(I)}(k\rho) &=&\left[\hat{\textbf{e}}_{\rho}\frac{in}{\rho}Z_{n}^{(I)}
(k\rho)-\hat{\textbf{e}}_{\varphi }k Z_{n}^{(I)^{\prime }}(k\rho)\right]e^{in\varphi },  \\
\mathbf{N}_{n}^{(I)}(k\rho) &=&\hat{\textbf{e}}_{z}k Z_{n}^{(I)}(k\rho)e^{in\varphi },
\end{eqnarray}
\end{subequations}
where $n$ is an integer;
$\rho$ is the radial distance and $\varphi$ is the azimuth angle;  $k$
is the wavevector; and
$Z_{n}^{(I)}$ is the solution of the Bessel equation with $I=1$ representing
the Bessel function of the first kind and $I=3$ the Hankel function of the first
kind.

We now consider a circular cylinder of TI with radius $r$ and
optical constants $\varepsilon $ and $\mu $ in air, schematically shown in Fig. 1
of the Letter. It should be mentioned that in our Letter we define a scattering angle
$\phi$ ($=\pi/2-\varphi$) and EM waves are incident in the $y$ direction.
The wavevector of incident waves is
$k=\omega /c$.
By using the vector cylindrical harmonics, incident and
scattered electric fields in air, and the internal electric field inside
 the TI cylinder can be written as
\begin{subequations}
\begin{eqnarray}
\mathbf{E}_{\text{inc}} &=&\overset{\infty }{\underset{n=-\infty }{\sum }}%
-E_{n}\left[ie_{i\perp}\mathbf{M}_{n}^{(1)}(k\rho)+e_{i\parallel}\mathbf{N}%
_{n}^{(1)}(k\rho)\right], \nonumber \\
\mathbf{E}_{\text{sca}} &=&\overset{\infty }{\underset{n=-\infty }{\sum }}%
E_{n}\left[ia_{n}\mathbf{M}_{n}^{(3)}(k\rho)+b_{n}\mathbf{N}_{n}^{(3)}(k\rho)\right],
\\
\mathbf{E}_{\text{int}} &=&\overset{\infty }{\underset{n=-\infty }{\sum }}%
E_{n}\left[c_{n}\mathbf{M}_{n}^{(1)}(k_{c}\rho)+d_{n}\mathbf{N}_{n}^{(1)}(k_{c}\rho)\right]
\nonumber,
\end{eqnarray}
\end{subequations}
where $a_{n}$ and $b_{n}$ are scattering coefficients;
$k_{c}=\sqrt{\varepsilon \mu }\omega /c$ is the wavevector of the TI cylinder;
$E_{n}=E_{0} /k$ with $E_{0}$ being the electric-field amplitude of the incident wave;
and $e_{i\perp}$ and $e_{i\parallel}$ are the polarization-vector components of incident
waves with  $e_{i\perp}=1$ and $e_{i\parallel}=0$ standing for the TE polarization
(with the electric field perpendicular to the cylinder) and $e_{i\perp}=0$ and $e_{i\parallel}=1$
for the TM polarization (with the electric field parallel to the cylinder).

The boundary conditions at the surface of the TI cylinder $\rho=r$ reads
\begin{subequations}
\begin{eqnarray}
(\mathbf{E}_{\text{inc}}+\mathbf{E}_{\text{sca}}-\mathbf{E}_{\text{int}})\times {\hat{\mathbf{e}}}_{\rho} &=&(\mathbf{H}_{\text{inc}}+\mathbf{H}_{\text{sca}}-\mathbf{H}_{int})\times
{\hat{\mathbf{e}}}_{\rho}=0, \\
(\mathbf{B}_{inc}+\mathbf{B}_{\text{sca}}-\mathbf{B}_{\text{int}})\cdot {\hat{\mathbf{e}}}_{\rho} &=&(\mathbf{D}_{\text{inc}}+\mathbf{D}_{\text{sca}}-\mathbf{D}_{\text{int}})\cdot
{\hat{\mathbf{e}}}_{\rho}=0,
\end{eqnarray}
\end{subequations}
and the constitutive relations for TIs are given by \cite{qi:08}
\begin{subequations}
\begin{eqnarray}
\mathbf{D} &=&\mathbf{\varepsilon E}- \overline{\alpha} \mathbf{H}, \\
\mathbf{H} &=&\mathbf{B}/\mu +\overline{\alpha} \mathbf{E}.
\end{eqnarray}
\end{subequations}
Here, $\overline{\alpha }=\alpha\theta /\pi$, where $\alpha$ is the fine structure
constant and $\theta=(2p+1)\pi$ with $p$ an integer is the axion angle, a quantized angular variable
to characterize the TME effect in TIs.
With the above boundary conditions and the constitutive relations, the scattering coefficients
can be found as
\begin{subequations}
\begin{eqnarray}
a_{n}&=&\frac{e_{i\perp}A_{n}D_{n}+\overline{\alpha }f_{n}}{A_{n}B_{n}+%
\overline{\alpha }^{2}t_{n}},\\
b_{n}&=&\frac{e_{i\parallel}B_{n}C_{n}+\overline{%
\alpha }g_{n}}{A_{n}B_{n}+\overline{\alpha }^{2}t_{n}}.
\end{eqnarray}
\end{subequations}
Here, $A_{n}$ ,$B_{n}$, $C_{n}$, and $D_{n}$ are the coefficients
for conventional dielectric cylinders defined in Ref. \cite{boh:83}, given by
\begin{subequations}
\begin{eqnarray}
A_{n}(x) &=&mJ_{n}^{^{\prime
}}(mx)H_{n}^{(1)}(x)-J_{n}(mx)H_{n}^{(1)^{\prime }}(x), \\
B_{n}(x) &=&mJ_{n}(mx)H_{n}^{(1)^{\prime }}(x)-J_{n}^{^{\prime
}}(mx)H_{n}^{(1)}(x), \\
C_{n}(x) &=&mJ_{n}^{^{\prime }}(mx)J_{n}(x)-J_{n}(mx)J_{n}^{^{\prime }}(x), \\
D_{n}(x) &=&mJ_{n}(mx)J_{n}^{^{\prime }}(x)-J_{n}^{^{\prime }}(mx)J_{n}(x),
\end{eqnarray}
\end{subequations}
where $x=kr$ is the size parameter and $m=\sqrt{\varepsilon \mu }$.
The coefficients $f_{n}$, $g_{n}$, and $t_{n}$ appear only in TI cylinders
due to the TME effect, given by
\begin{subequations}
\begin{eqnarray}
f_{n}(x) &=&J_{n}(mx)J_{n}^{^{\prime }}(mx)\left[ -e_{i\parallel}\frac{2i}{%
\pi x}+\overline{\alpha }e_{i\perp}J_{n}^{^{\prime
}}(x)H_{n}^{(1)}(x)\right], \\
g_{n}(x) &=&J_{n}(mx)J_{n}^{^{\prime }}(mx)\left[ -e_{i\perp}\frac{2i}{\pi x%
}+\overline{\alpha }e_{i\parallel}J_{n}(x)H_{n}^{(1)^{\prime }}(x)\right], \\
t_{n}(x) &=&J_{n}(mx)J_{n}^{^{\prime }}(mx)H_{n}^{(1)}(x)H_{n}^{(1)^{\prime
}}(x).
\end{eqnarray}
\end{subequations}

The physical meaning of the scattering coefficients $a_{n}$ and $b_{n}$ lies in
the fact that they are associated with
the electric and magnetic multipoles of order $n$, respectively. It can be verified that
$a_{n}=a_{-n}$ and $b_{n}=b_{-n}$, similar to those in conventional dielectric
cylinders \cite{boh:83,tsa:00}.
For conventional dielectric cylinders ($\theta =0$ or $\overline{\alpha }=0$),
the scattering coefficients $a_{n}$ and $b_{n}$ are reduced to be the ones
given in Ref. \cite{boh:83,tsa:00} as expected. Note that for incident EM waves there are
two independent polarizations: TE and TM. As a result, there exist two sets of
scattering coefficients, namely, \{$a_{n,\text{TE}}$,\ $b_{n,\text{TE}}$\} and
\{$a_{n,\text{TM}}$,\ $b_{n,\text{TM}}$\}.

In the Rayleigh limit, i.e., $x \ll 1$ and $mx\ll 1$, the scattering coefficients can
be expanded by Tailor series. If only the terms up to the order of $x^{2}$ are kept,
for TE polarization $a_{n}$ and $b_{n}$ can be approximated as
\begin{subequations}
\begin{eqnarray}
a_{0,\text{TE}}&=& 0+O(x^{4}), \\
a_{1,\text{TE}}&=& -\frac{i\pi (2m^{2}-2+\overline{\alpha }^{2})}
{4(2m^{2}+2+\overline{\alpha }^{2})}x^{2}+O(x^{4}),\\
b_{0,\text{TE}}&=& \frac{i\pi \overline{\alpha }}{4}x^{2}+O(x^{4}),\\
b_{1,\text{TE}}&=& -\frac{i\pi \overline{\alpha }}{2(2m^{2}+2+
\overline{\alpha }^{2})}x^{2}+O(x^{4}).
\end{eqnarray}
\end{subequations}
Obviously, in the Rayleigh limit only the electric dipole, and magnetic monopole
and dipole involve in the scattering for TE incident waves. The electric monopole
cannot be excited. In conventional dielectric cylinders, however, only the
electric dipole plays a role in scattering.

For TM incident EM waves, the scattering coefficients can be approximated in
the Rayleigh limit as
\begin{subequations}
\begin{eqnarray}
a_{0,\text{TM}}&=& b_{0,\text{TE}}, \\
a_{1,\text{TM}}&=& b_{1,\text{TE}}, \\
b_{0,\text{TM}}&=& -\frac{i\pi }{4}%
(m^{2}-1+\overline{\alpha }^{2})x^{2}+O(x^{4}), \\
b_{1,\text{TM}}&=& \frac{i\pi \overline{\alpha }^{2}}{%
4(2m^{2}+2+\overline{\alpha }^{2})}x^{2}+O(x^{4}).
\end{eqnarray}
\end{subequations}
In additional to the electric dipole, and magnetic monopole and dipole,
the electric monopole can be excited for TM incident waves.

\setcounter{equation}{0}

\subsection{Appendix B: Evaluation of the size effect }
\renewcommand{\theequation}{B.\arabic{equation}}

\begin{figure}[H]
\centerline{\includegraphics[width=8cm]{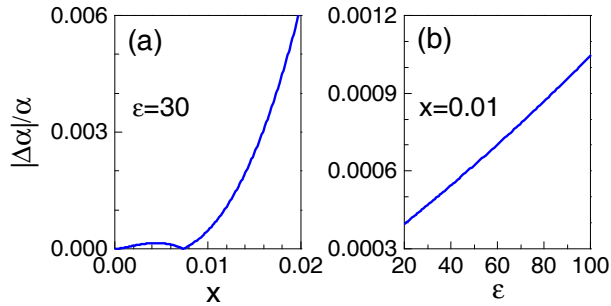}}
\caption{ (a) $|\Delta\alpha|/\alpha$  as a function of $x$ for a TI cylinder
with $\varepsilon=30$ and $\theta=\pi$.
(b) $|\Delta\alpha|/\alpha$ as a function of the dielectric constant $\varepsilon$
of a TI cylinder with $x=0.01$ and $\theta=\pi$.}
\label{fig:a1}
\end{figure}

To estimate the accuracy of the optical measurement of the fine structure constant
$\alpha$ , we introduce a deviation function
\begin{equation}
\Delta\alpha=\alpha(x,\varepsilon)-\alpha,
\end{equation}
where $\alpha(x,\varepsilon)$ is no longer a constant but is a function, for example,
defined by Eq. (6) in the two-angle measurement in the Letter.  In Eq. (6), the quantity
$R(\phi)=|E_{s\parallel}/E_{s\perp}|$ is calculated by the rigorous Mie theory without
taking the Rayleigh limit. In Fig.~\ref{fig:a1}(a), two scattering angles 
$\phi_{1,2}=90^{\circ}\pm 9^{\circ}$ are chosen, and the deviation,
$|\Delta\alpha/\alpha|$ as a function of the size parameter $x$ with
$\varepsilon$=30 fixed is shown. Obviously, $|\Delta\alpha/\alpha|$ is in the order of
$10^{-4}$ for $x<0.01$. Figure ~\ref{fig:a1}(b) shows $|\Delta\alpha/\alpha|$
as a function of the dielectric constant with $x=0.01$, which is always in the order of
$10^{-4}$  as  $\varepsilon$ varies from 20 to 80.

\end{appendix}

\end{document}